\documentclass[10pt]{iopart}
\usepackage{amssymb}
\usepackage{graphicx}

\newcommand{\beq}[0]{\begin{equation}}
\newcommand{\eeq}[0]{\end{equation}}

\newcommand{\la}{\langle}
\newcommand{\ep}{\varepsilon}
\newcommand{\ra}{\rangle}
\newcommand{\ds}{\displaystyle}
\newcommand{\avh}{\la H_1 \ra}
\newcommand{\pa}{\partial}
\newcommand{\w}{\omega}
\newcommand{\ud}{{\mathrm d}}

\begin{document}
\title{Existence and stability of multisite breathers in honeycomb and hexagonal lattices}

\author{V. Koukouloyannis$^{1, 2}$, P.G. Kevrekidis$^{3}$, K.J.H. Law$^3$, \\I. Kourakis$^4$, D.J. Frantzeskakis$^5$}
\address{$^{1}$Department of Physics, Section of Astrophysics, Astronomy and Mechanics, Aristotle University of Thessaloniki, 54124 Thessaloniki, Greece}
\address{$^{2}$Department of Civil Engineering, Technological Educational Institute of Serres, 62124 Serres, Greece}
\address{$^3$Department of Mathematics and Statistics, University of Massachusetts, Amherst MA 01003-4515}
\address{$^4$Centre for Plasma Physics, Queen's University Belfast, BT7 1 NN Northern Ireland, UK}
\address{$^5$Department of Physics, University of Athens, Panepistimiopolis, Zografos, Athens 15784, Greece}


\pacs{63.20.Pw, 63.20.Ry, 05.45.Yv}

\begin{abstract}
We study the existence and stability of multisite discrete breathers in two prototypical non-square Klein-Gordon lattices, namely
a honeycomb and a hexagonal one. In the honeycomb case we consider six-site configurations and find that for soft potential and positive coupling the
out-of-phase breather configuration and the charge-two vortex breather are linearly stable, while the in-phase and charge-one vortex states
are unstable. In the hexagonal lattice, we first consider
three-site  configurations. In the case of soft potential and positive coupling, the in-phase configuration
is unstable and the charge-one vortex is linearly stable.  The
out-of-phase configuration here is found to always be linearly
unstable.
 We then turn to six-site configurations in the hexagonal lattice. The stability results in this case are the same as in the
  six-site configurations in the honeycomb lattice. For all configurations in
both lattices, the stability results are reversed in the
setting of either hard potential or negative coupling.
The study is complemented by numerical simulations which
are in very good agreement with the theoretical predictions.
Since neither the form of the on-site potential nor the sign of the coupling parameter involved have been prescribed, this description can accommodate inverse-dispersive systems (e.g., supporting backward waves) such as transverse dust-lattice oscillations in dusty plasma (Debye) crystals or analogous modes in molecular chains.
\end{abstract}

\maketitle

\section{Introduction}
Over the past decade, there has been a considerable increase of
interest in the study of discrete or quasi-discrete systems in a wide
range of areas in physics. Among the numerous themes of intense
theoretical and experimental interest, one can
refer to the DNA double-strand dynamics in biophysics \cite{peyrard},
coupled waveguide arrays and photorefractive crystals
in nonlinear optics \cite{photon,moti3,review_opt}, breathing oscillations
in micromechanical cantilever arrays \cite{sievers},
Bose-Einstein condensates in optical lattices in atomic
physics \cite{morsch1},
granular crystals \cite{sen08}, and so on.

These studies have been conducted predominantly in one-dimensional (1D)
systems or in higher dimensions, but chiefly in the context of square lattices.
In the latter,
a wide variety of novel and interesting phenomena
have been revealed, both theoretically and
experimentally.
Pertinent examples include, among others, the prediction and observation of
dipole \cite{dip}, and necklace \cite{neck} solitons, discrete vortices
\cite{vortex1,vortex2}, rotary solitons \cite{kartashov,rings}, higher-order
Bloch modes \cite{neshev2} and gap vortices \cite{motihigher},
as well as two-dimensional Bloch oscillations and Landau-Zener tunneling \cite{zener}.

More recently, the dynamics of non-square lattices have become
a focal point of interest, both in the context of periodic
photonic structures~\cite{honey,rosberg2,gaid,tja2}, and in that
of quasi-crystalline \cite{motinature1} or
completely disordered lattices \cite{motinature2}.
While most of the above works had a view towards applications
based on photorefractive crystals, there exist
many other applications where such non-square lattices may be relevant.
In particular, for the
hexagonal and honeycomb
lattices considered herein,
we note that they have already been showcased in recent experiments in
two-dimensional waveguide arrays (e.g., in glass) \cite{szameit},
in optical lattices acting on Bose-Einstein condensates \cite{klaus},
as well as in Debye crystals formed in dusty plasmas \cite{yannis,koukkour2}.
Interestingly, in the latter case (dusty plasma crystals), the horizontal propagation of transverse (off-plane) vibrations gives rise to an inverse-optic dust-lattice mode (a backward wave) \cite{Vlad1, SVbook} which is described by a Klein-Gordon-type equation like the one we shall focus on below, yet upon formally considering a negative coupling coefficient (``spring constant"). In this case, nonlinearity is provided by the (anharmonic) plasma sheath potential, while discreteness is expressed by the (small value of the) ratio of a characteristic coupling frequency (related to electrostatic Debye interactions). Remarkably, both ab initio theoretical considerations and experimental findings suggest that the sheath on-site potential is intrinsically anharmonic and in fact strongly asymmetric, so that existing theories involving even (symmetric) polynomial functions for the on-site potential do not apply in this case. Further contribution to nonlinearity is furnished by Debye-type (screened Coulomb) electrostatic inter-dust-particle interactions, yet the associated anharmonicity is of lesser order of magnitude for the transverse mode and can be neglected.
Details on dusty plasma modelling can be found, e.g., in \cite{IKIJBC, PKSbook}, while the experimental setting is described in \cite{SVbook, GEMreview}.

The above experimental developments
have
prompted further theoretical work towards the goal of understanding
the structures that may emerge in such non-square lattices and
their corresponding stability. An example of this type is the
recent work of \cite{lawetal}, where this analysis was performed
in the framework of the discrete nonlinear Schr{\"o}dinger (DNLS) equation,
a prototypical model widely perceived as relevant to optical systems.

The aim of the present work, motivated  by models
of dusty plasma lattices \cite{yannis,koukkour2}, is to extend the
considerations in \cite{lawetal} to nonlinear Klein-Gordon lattices of the hexagonal and honeycomb type.
We stress the fact that no assumption will be made on the (anharmonic) substrate potential or the sign of the coupling parameter involved in the description. In particular, inverse-dispersive systems (supporting backward propagating waves in the linear limit) such as transverse oscillations in dusty plasma crystals are straightforward to accommodate in this description (by reverting, say, the sign of the ``spring" coupling coefficient to negative).
Within this framework, we offer a systematic analysis of the type
of solutions that arise in six-site contours in honeycomb lattices
and three-site as well as six-site contours in hexagonal lattices
(for the latter lattice, see also  earlier work
in Refs. \cite{koukkour2,koukmac}).
In the six-site contours, we illustrate that solutions with higher topological charge $S$
(namely, $S=2$)
are more dynamically robust
than lower topological charge ones (namely, $S=1$).
This is true for soft nonlinearities and positive coupling (or hard nonlinearities and negative coupling),
while the results are reversed if either the nature of the nonlinearity
or the sign of the coupling are changed.
It is interesting to note that similar findings have been reported
not only in DNLS type chains  \cite{lawetal}
but also in continuum photorefractive
crystals \cite{alexlaw} (where they
have recently been confirmed experimentally
\cite{berndlaw}). Furthermore, in-phase
and out-of-phase
structures are also examined. The former are found
to be unstable, while the latter are potentially linearly stable.
In the three-site, hexagonal configuration, both in- and out-of-phase
structures are unstable, while the vortex one may be stable. Again, this is true
for soft nonlinearities and positive coupling (or hard nonlinearities and negative coupling),
while all the results, besides the out-of-phase case, are reversed if either the nature of the nonlinearity
or the sign of the coupling are changed. Finally, in the six-site configurations in the hexagonal lattice, the stability results are the same as in the case of six-site configurations in the honeycomb lattice.

Our presentation is structured as follows.
In section II, we analyze the six-site honeycomb contour, while
in section III, we briefly review the findings for the three-site hexagonal
contour. Section IV contains the comparison with
numerical results.
Finally, section V contains a summary of our
findings and a brief discussion of future directions.

\section{Existence of multisite breathers in a honeycomb Klein-Gordon lattice}

\begin{figure}[tbp]
	\centering
		\includegraphics[width=8cm]{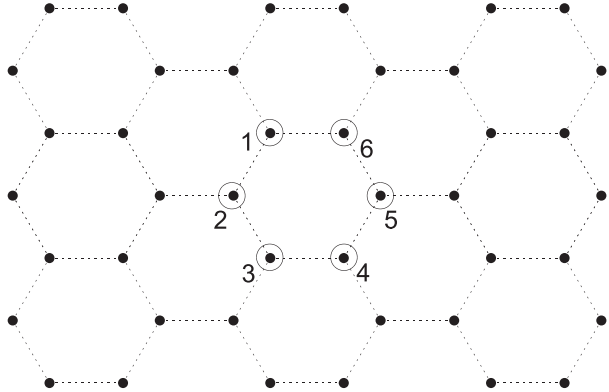}
\caption{The honeycomb lattice and associated six-site contours}	\label{lattice_fig}
\end{figure}

We consider a honeycomb lattice
(see the protoypical example of figure \ref{lattice_fig}) with on-site
potential and linear nearest neighbor interaction.
This system is described by a Klein-Gordon Hamiltonian of the form
\begin{equation}
H=H_0+\ep H_1=\sum_{i\in \mathbb{S}} \frac{p_i^2}{2}+V(x_i)+\frac{\ep}{2}\sum_{i,j\in \mathbb{G}} (x_j-x_i)^2,
\label{hamilt}
\end{equation}
where $\mathbb{S}$ is the set of all the oscillators and $\mathbb{G}$ is the set of neighboring pairs of oscillators.
In the anti-continuum (AC) limit \cite{macaub} ($\ep=0$) we consider the six encircled oscillators of figure \ref{lattice_fig}
as moving in periodic orbits with the same period $\w_i=\w$, while the rest of the oscillators lie at equilibrium $(x_i, p_i)=(0, 0)$.
This periodic and trivially localized motion is continued, for $\ep\neq0$ small enough, to provide multi-site breathers,
if the phase differences between successive oscillators $\phi_i$ are such that they correspond to critical points of the
effective Hamiltonian $H^{\mathrm{eff}}$
\cite{ahnmacsep}. To leading-order of approximation,
the effective Hamiltonian becomes
$$H^{\mathrm{eff}}_1(I_i,\phi_i,A)=H_0(I_i,A)+\ep\la H_1\ra(I_i,A,\phi_i),$$
where $I_i, A, \phi_i$ are canonical variables which are defined by the transformation
\beq\begin{array}{lll}
\vartheta=w_1, & &A=\ds\sum_{j=1}^6J_j\\[8pt]
\phi_i=w_{i+1}-w_i,& &I_i=\ds\sum_{j=i+1}^6 J_j
\qquad i=1, \ldots 5
\\
\end{array}
\label{transformation}
\eeq
with $J_i$, $w_i$
being the action-angle variables. The average value of $H_1$, namely $\ds\avh=\frac{1}{T}\oint H_1\ud t$, is calculated along the unperturbed orbit.
Note that, since the motion of a single oscillator for $\ep=0$ can be described by $J_i=\mathrm{const.}$ and $w_i=\w t+w_{0i}$,
the variables $\phi_i$ can be considered as the phase differences between two successive oscillators. Since we consider six oscillators in
the
AC limit and $\phi_6=-\sum_{i=1}^5\phi_i$, we get only five independent $\phi_i$. Then, the configurations of the
AC limit that will be continued to provide the multisite breathers correspond to simple roots of the system of equations
\beq
\frac{\pa\avh}{\pa\phi_i}=0,
\quad i=1\ldots5
\label{cond}
\eeq
provided, of course, the non-resonance of the frequency of the breather $\w_b$ with the frequency $\w_b$ of the system's linear spectrum, namely
$k\w_b\neq\w_p$,
and the anharmonicity of the single uncoupled oscillator, i.e.\, $\pa\w/\pa J\neq0$. This result is in accordance with the findings
of \cite{koukicht1} (see also \cite{koukmac}).

In order to calculate $\avh$ we use the fact that the motion of each oscillator in the
AC limit can be described by a cosine Fourier series due to the time-reversibility [viz. $x(-t)=x(t)$, $p(-t)=-p(t)$],
\beq x_i(t)=\sum_{n=0}^{\infty}A_n(J_i)\cos(nw_i)
\label{fourier}
\eeq
where $(J,w)$ are the action-angle variables for the specific oscillator. Using this fact, and the canonical transformation
(\ref{transformation}), $\avh$ becomes
$$\la H_1\ra=-\frac{1}{2}\sum_{n=1}^{\infty}A_n^2\left\{\cos(n\phi_1)+\cos(n\phi_2)+\cos(n\phi_3)
+\cos(n\phi_4)+\cos(n\phi_5)+\cos[n(\phi_1+\phi_2+\phi_3+\phi_4+\phi_5)] \right\}$$
So, condition (\ref{cond}) becomes
$$\sum_{n=1}^{\infty}nA_n^2\left\{\sin(n\phi_i)+\sin[n(\phi_1+\phi_2+\phi_3+\phi_4+\phi_5)] \right\}=0,
\quad i=1\ldots5,
$$
which leads to two groups of solutions. The first one is
$$\phi_i=0, \qquad {\rm or} \qquad \pi,
\qquad i=1,\ldots,5,
$$
which determines the time-reversible solutions according to the
terminology of \cite{macaub}, while the second kind of solutions
$$\phi_i=S \frac{\pi}{3}\qquad i=1,
\ldots,5\quad \mathrm{and}
\quad S\in\{1,2\}
$$
determines the non-time-reversible solutions, i.e., vortex breather solutions of topological charge $S$.
This kind of solutions can be distinguished from the previous one because it possesses a nonzero energy flux \cite{creaub}.
The form and the time evolution over a period of the above mentioned motions can be found in \cite{video_link}.

The stability of the above mentioned solutions is determined through the corresponding Floquet multipliers
$\lambda_i$ (see e.g. \cite{aubrev}). The {\it characteristic exponents} of the breather are defined as
$\lambda_i=e^{\sigma_iT}$. The nonzero characteristic exponents of the central oscillators are given as
eigenvalues of the stability matrix $E=\Omega D^2H^\mathrm{eff}$ where $\Omega=\left(\begin{array}{cc}O&-I\\I&O\end{array}\right)$
is the symplectic structrure matrix and $O, I$ are the $5\times5$ zero and identity matrices respectively. The leading order approximation of $E$ is given by
\beq E_1=\left(\begin{array}{c|c}
-\ep\ds\frac{\pa^2\avh}{\pa\phi_i\pa I_j}&-\ep\ds\frac{\pa^2\avh}{\pa\phi_i\pa\phi_j}\\[10pt]
\hline\\[-8pt]
\ds\frac{\pa^2H_0}{\pa I_i I_j}+\ds\ep\frac{\pa^2\avh}{\pa I_i\pa I_j}&\ds\ep\frac{\pa^2\avh}{\pa\phi_j\pa I_i}
\end{array}\right).
\label{e1}
\eeq
If all the eigenvalues of matrix $E$ lie on the imaginary axis, then the
corresponding breather is linearly stable. If the eigenvalues of $E_1$ are
simple to leading-order in $\ep$ and lie on the imaginary axis, then the eigenvalues of $E$ will also lie in the imaginary axis,
due to continuity reasons. If the eigenvalues of $E_1$ are not simple in the leading-order, then the higher-order terms of the
approximation could push the eigenvalues of $E$ outside the imaginary axis and, thus, lead to complex instability.
However, this cannot happen if the symplectic signature of $E_1$ is definite \cite{mac2} and the breather remains stable for
$\ep$ small enough. This signature is definite if the quadratic form $x^TE_1x$ has the same sign for every vector $x\in\mathbb{R}^{10}$.

Note that, although the above conditions certify linear stability for small values of $\ep$, for higher values of $\ep$
a characteristic exponent of the central oscillators can collide with the linear spectrum causing instability through
a Hamiltonian-Hopf bifurcation, as it can be seen in some of our
numerical results below (see also \cite{aubrev}).

\subsection{The four configurations under consideration}\label{e1calc}
We consider four representative configurations of the
AC limit which lead to the corresponding multisite breathers. For all of the
ones we consider, it is true that $\w_i=\w\Leftrightarrow J_i=J$ and $\phi_i=\phi$, $i=1\ldots 5$.
More precisely, we consider the {\it in-phase} configuration, $\phi=0$, the {\it out-of-phase} one, $\phi=\pi$, the charge-one vortex
breather $\phi=\pi/3$ and the charge-two vortex breather $\phi=2\pi/3$. Using (\ref{transformation}), we get
for the various elements of matrix $E_1$,
$$\begin{array}{cc}\ds\frac{\pa^2H_0}{\pa I_i \pa I_{j}}=&\left\{\begin{array}{cccl}
\ds2\frac{\pa^2H_0}{\pa J^2}&=&\ds2\frac{\pa\w}{\pa  J}, &j=i\\[10pt]
\ds-\frac{\pa^2H_0}{\pa J^2}&=&\ds-\frac{\pa\w}{\pa J}, &j=i+1\\[8pt]
\ds 0& & &\mathrm{otherwise}
\end{array}\right.\end{array}
$$

$$\begin{array}{cc}\ds\frac{\pa^2\avh}{\pa \phi_i\pa \phi_j}=&\left\{\begin{array}{ll}
\ds 2f(\phi), &j=i\\[8pt]
\ds  f(\phi), &j\neq i
\end{array}\right.\end{array},\quad \mathrm{with}\quad f(\phi)=\frac{1}{2}\sum_{n=1}^{\infty}n^2A_n^2\cos(n\phi)$$

$$\begin{array}{cc}\ds\frac{\pa^2\avh}{\pa I_i\pa I_{j}}=&\left\{\begin{array}{ll}
\ds 2g_1-3g_2, &j=i\\[8pt]
\ds -2g_1-2g_2, &j=i+1\\[8pt]
\ds g_1, &j=i+2\\[8pt]
\end{array}\right.\end{array} ,\  \mathrm{with}\quad
\begin{array}{l}
\ds g_1=\frac{1}{2}\sum_{n=1}^\infty\left(\frac{\pa A_n}{\pa J}\right)^2\cos(n\phi)\\[13pt]
\ds g_2=\frac{1}{2}\sum_{n=1}^\infty\frac{\pa^2 A_n}{\pa J^2}\cos(n\phi)
\end{array}$$

$$\begin{array}{cc}\ds\frac{\pa^2\avh}{\pa I_i\pa \phi_j}=&\left\{\begin{array}{ll}
\ds k, &j=i+1\\[8pt]
\ds  0, &j\neq i+1
\end{array}\right.\end{array} ,\ \mathrm{with}\quad k=\frac{1}{2}\sum_{n=1}^\infty n\frac{\pa A_n}{\pa J}A_n \sin(n\phi)$$
Below, we examine separately each of the four principal configurations under investigation.
Up to leading-order of approximation, the eigenvalues of the stability matrix $E_1$ correspond to the ${\cal O}(\sqrt{\ep})$ approximation of the characteristic exponents, and they are given, for all the configurations under consideration, by the expressions:
\beq
\sigma_{\pm1,\pm2}=\pm\sqrt{-\ep\frac{\pa \w}{\pa J}f},\quad
\sigma_{\pm3\pm4}=\pm\sqrt{3}\sqrt{-\ep\frac{\pa \w}{\pa J}f},\quad\sigma_{\pm5}=\pm2\sqrt{-\ep\frac{\pa \w}{\pa J}f} \, . \quad
\label{s_hon_in}
\eeq

It may be added for rigor that we have chosen to keep the sign of the coupling strength $\ep$ arbitrary, bearing in mind
that inverse dispersive systems, such as transverse dust-lattice vibrations (briefly discussed above),  require negative values of $\ep$ to be considered (in contrast with the ordinary Klein-Gordon formulation). Thus, $\ep$ may be either negative or positive throughout this text, unless otherwise stated.

\subsubsection{The in-phase multibreather $\phi=0$. \label{in}}

We first examine the in-phase ($\phi=0$) configuration. In this case we get $\frac{\pa^2\avh}{\pa I_i\pa \phi_j}=0$,
which simplifies the calculations of the characteristic exponents $\sigma_i$ in equation (\ref{s_hon_in}). Since for continuous periodic functions
the size of Fourier coefficients $A_n$ is exponentially decreasing
as a function of $n$, $f$ converges -- in this case to a positive number:
$$f=f(0)=\frac{1}{2}\sum_{n=1}^{\infty}n^2A_n^2>0.$$
Hence, the nature of the exponents (and the linear stability)
hinges on the product $\ep\frac{\pa \w}{\pa J}$. If it is negative
(as is the case for soft nonlinearities and positive couplings,
or for hard nonlinearities and negative couplings), the
exponents are real and the corresponding breather is unstable. If
$\ep\frac{\pa \w}{\pa J}>0$ we have to check also the corresponding
symplectic signature
to ensure that the exponents will remain in the imaginary axis.
It can be proved (but the calculations are rather lengthy) that this
signature is definite if $\ep\frac{\pa \w}{\pa J}>0$. So, in this case
the resulting breather is linearly stable for $\ep$ small enough in order to avoid collisions with the linear spectrum.

\subsubsection{The out-of-phase multibreather, $\phi=\pi$.}\label{out}
For $\phi=\pi$ the stability matrix $E_1$ and its eigenvalues ($\sigma_i$) are the same as before but in this case it is
$$f=f(\pi)=\frac{1}{2}\sum_{n=1}^{\infty}n^2A_n^2(-1)^n<0.$$
Due to the Hamiltonian nature of the single oscillator it is  $A_1>4A_2$ and since the absolute value
of the terms is exponentially decreasing, $f$ converges to a negative number.
Following similar
arguments as in the previous section we conclude that
for $\ep\frac{\pa \w}{\pa J}<0$ (under the physical conditions
discussed in the previous subsection), the breather is linearly stable,
while it is unstable otherwise.

\subsubsection{The vortex breather of charge $S=1$, $\phi=\pi/3$.}\label{s1}
%
In the case of the vortex with charge $S=1$ ($\phi=\pi/3$)
the terms $\ds\frac{\pa^2\avh}{\pa I_i\pa \phi_j}$ are not identically zero, but the corresponding characteristic exponents are still, to leading-order of approximation, given by equation (\ref{s_hon_in}). For the polynomial potentials typically used,
$f$ converges to a positive number:
$$f=f(\frac{\pi}{3})=\frac{1}{2}\sum_{n=1}^{\infty}n^2A_n^2\cos\left(\frac{n\pi}{3}\right)>0.$$
%
Thus, following the arguments of the previous sections for $\ep\frac{\pa \w}{\pa J}<0$ it turns out that
the vortex breather is linearly unstable, while for $\ep\frac{\pa \w}{\pa J}>0$ the breather is linearly stable.

\subsubsection{The vortex breather of charge $S=2$.}\label{s2}
The stability matrix $E_1$ and its eigenvalues ($\sigma_i$) are the same as before, but since for $\phi=2\pi/3$,
it is found that
$$f=f(\frac{2\pi}{3})=\frac{1}{2}\sum_{n=1}^{\infty}n^2A_n^2(-1)^n\cos\left(\frac{n\pi}{3}\right)<0.$$
Thus, we conclude that the corresponding breather is linearly stable if $\ep\frac{\pa \w}{\pa J}<0$, while otherwise it is unstable.
Note that the symplectic signature arguments mentioned in subsection \ref{in} are still valid.

\section{Existence and stability of 3-site and 6-site breathers in a Hexagonal Klein Gordon Lattice}

\begin{figure}[htb]
	\centering
		\includegraphics[width=8cm]{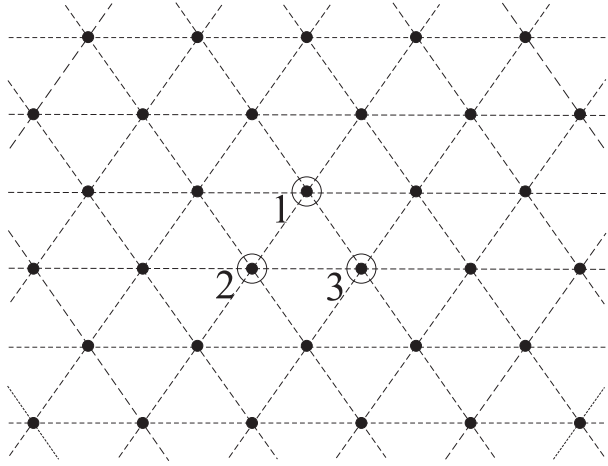}
\caption{The hexagonal lattice and associated contours.}	\label{hex_lattice_fig}
\end{figure}

Let us start by briefly reviewing the theory of Refs. \cite{koukkour2,koukmac}, as
a preamble to the systematic comparison with numerical results in the following section.
The Klein-Gordon Hamiltonian is of the form
of equation (\ref{hamilt}),
but now each site has six neighbors instead of three as before.

We consider
an AC limit of three oscillators,
as per the contour shown in figure ~\ref{hex_lattice_fig}. In this case,
there exist only two independent $\phi_i$'s since
$\phi_3=w_1-w_2=-\phi_1-\phi_2$.
The persistence conditions for 3-site breathers in these lattices are
$$\sum_{n=1}^{\infty}nA_n^2\left\{\sin
(n\phi_i)+\sin\left[ n(\phi_1+\phi_2)\right]\right\}=0 \quad i=1,2.$$
Seeking zeros of the curly bracket in the above expression,
we will consider the {\it in-phase} configuration with
$\phi_i=0$,
the {\it out-of-phase} one with
$\phi_i=\pi$,
and the {\it vortex breather} with
$\phi_i=\frac{2\pi}{3}$. The form and time evolution over a period of these motions can also be found in \cite{video_link}.
To leading-order of approximation, the corresponding characteristic exponents are given by the following expressions:
for the in-phase configuration:
\beq
\sigma_{\pm1,\pm2}= \pm \sqrt{-3\ep\frac{\pa \w}{\pa J}f(0)},
\label{s_in}
\eeq
for the out-of-phase configuration:
\beq\sigma_{\pm1}=\pm\sqrt{-\ep\frac{\pa\w}{\pa J} \, [2f(0)+f(\pi)]},\qquad
\sigma_{\pm2}=\pm\sqrt{-3\ep\frac{\pa\w}{\pa J}\, f(\pi)},
\label{s_out}
\eeq
and, finally, for the vortex configuration:
\beq
\sigma_{\pm1,\pm2}=\pm\sqrt{-3\ep\frac{\ud\w}{\ud J}\, f(2\pi/3)}, \label{s_vortex} \,
\eeq
where the function $f(\phi)$ is defined as in the previous section.

In the case of the six-site configurations, the characteristic exponents results are the same as in the honeycomb lattice and are given by equations (\ref{s_hon_in}). This is because our theory is a first order one, while the influence of the extra sites in this case would be visible in the higher order of the expansion of the exponents.

\section{Numerical Results}

We perform a set of numerical computations in order to demonstrate the validity of our results.
Although the theoretical analysis above is completely general, in what
follows, we consider a particular choice of an on-site potential and
breather frequency. More specifically,
the potential under consideration is a quartic one, of the form,
$$V(x)=\frac{a}{2}x^2+\frac{b}{3}x^3+\frac{c}{4}x^4$$
with $a=1, b=-0.27, c=-0.03$. For this set of parameters $\frac{\pa \w}{\pa J}<0$.
We will consider the orbit with frequency $\w=7.43409$ which corresponds to amplitude of oscillation
$x_{max}=1.949275$; thus, $J=1.20306$ and $\frac{\pa\w}{\pa J}=-0.224556$.
For the same orbit we get $f(0)=1.423404$, $f(\pi)=-1.279544$, $f\left(\frac{\pi}{3}\right)=0.638983$ and $f\left(\frac{2\pi}{3}\right)=-0.710913$.

We point out that a positive value was considered in our numerical investigation below. Recalling that the value of $\ep$ may be either positive or negative (see discussion above), one should therefore keep in mind that the qualitative predictions (on breather stability) to follow are directly reversed if either the
sign of $\ep$ or the sign of $\frac{\pa \w}{\pa J}$ are
reversed (to negative/positive, respectively).
They obviously remain unchanged if \emph{both} quantities change sign.

\subsection{Honeycomb lattice}

\subsubsection{In-phase, 6-site breather.}
\begin{figure}[!ht]
	\centering
		\includegraphics{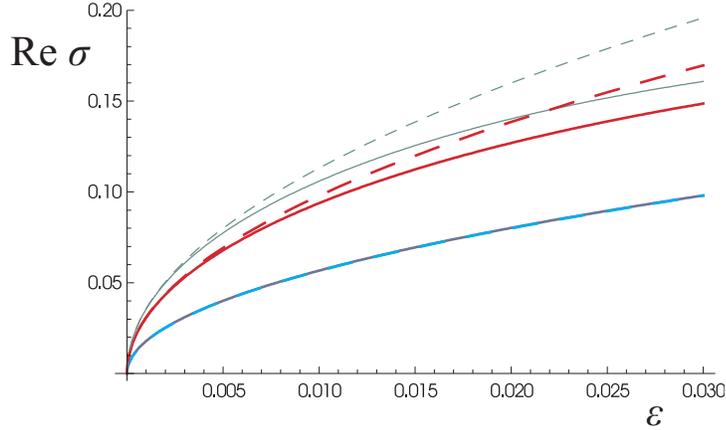}
	\caption{The real part of the characteristic exponents $\sigma_i$ for the
in-phase honeycomb configuration for increasing values of $\ep$. The solid lines
represent the values obtained numerically,
while the dashed ones represent the ${\cal O}(\sqrt{\ep})$ theoretical prediction.}
	\label{fig:eig_hon_in_re}
\end{figure}

First we consider the in-phase configuration $(\phi_i=\phi=0)$. As
mentioned in section \ref{in}, all the characteristic exponents $\sigma_i$ of the central oscillators are real to leading-order of approximation.
We calculate the breather for increasing values of $\ep$ and numerically compute the corresponding characteristic exponents.
In figure ~\ref{fig:eig_hon_in_re} we show the theoretically predicted values [cf. equation (\ref{s_hon_in})], depicted by
dashed lines, together with the ones calculated by the numerical simulation, depicted by solid lines.
In this figure, as well as in the following ones,
shown is only the positive $\sigma_i$.
It is observed that the theoretical and numerical branches almost coincide for small values of $\ep$, while for larger values of $\ep$,
where the higher-order terms of $\sigma_i$ become significant,
the lines start to deviate from each other.
We note also that, instead of having five branches
of numerical $\sigma_i$, we have only three. This happens because the branches of the $\sigma_{1,2}$ pair, as well as the $\sigma_{3,4}$ pair coincide
(these double pairs will be hereafter denoted by thicker lines), which means that the higher-order terms of the approximation also coincide.

\subsubsection{Out-of-phase, 6-site breather.}
\begin{figure}[!ht]
	\centering
		\includegraphics{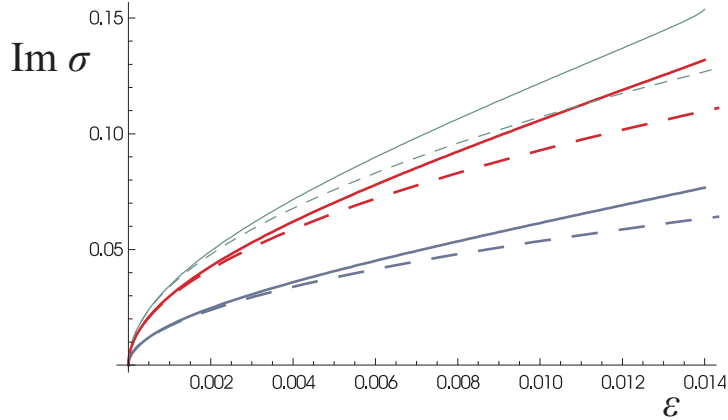}
	\caption{The imaginary part of the characteristic exponents
$\sigma_i$ for the out-of-phase honeycomb configuration for increasing values of $\ep$. The solid lines represent
the values obtained numerically, while the dashed ones represent the ${\cal O}(\sqrt{\ep})$ theoretical prediction.}
	\label{fig:eig_hon_out}
\end{figure}

The next configuration is the out-of-phase one, i.e.\, $\phi_i=\phi=\pi$. As it has already been discussed in section \ref{out},
this is a linearly stable configuration. Indeed, we have numerically
calculated the corresponding characteristic exponents for increasing
values of the coupling constant $\ep$ and
accordingly confirmed that all of them lie on the imaginary axis,
as it was expected. The calculated value
of the exponents, as well as the theoretical ${\cal O}(\sqrt{\ep})$ prediction, are shown in figure \ref{fig:eig_hon_out}.
We note again that, for small values of $\ep$, the theoretical (dashed) with the numerical (solid) curve coincide, and for
larger values of $\ep$, where the higher order terms become significant, they
start to diverge.

\subsubsection{ Charge $S=1$, 6-site vortex breather.}
\begin{figure}[t]
	\centering
		\includegraphics[width=6cm]{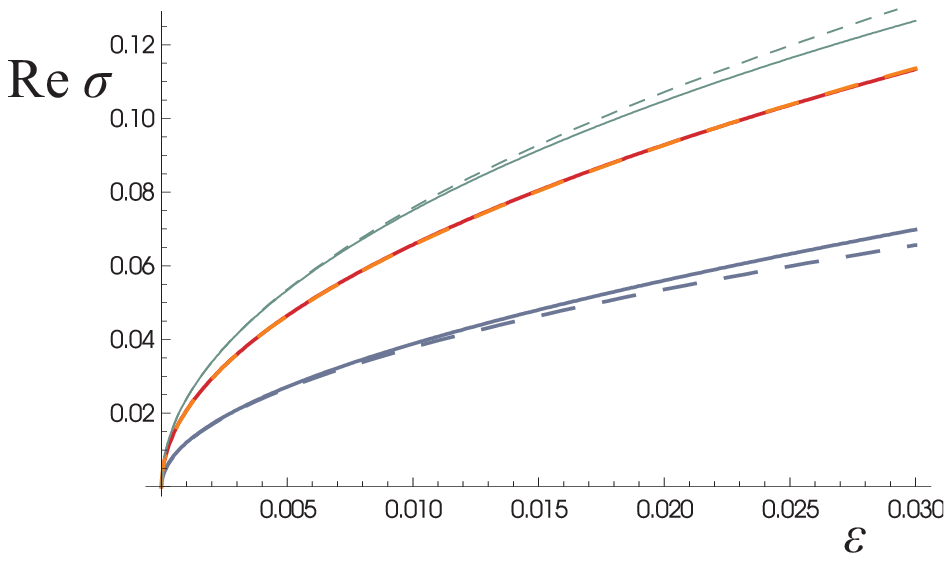} \ \ \includegraphics[width=6cm]{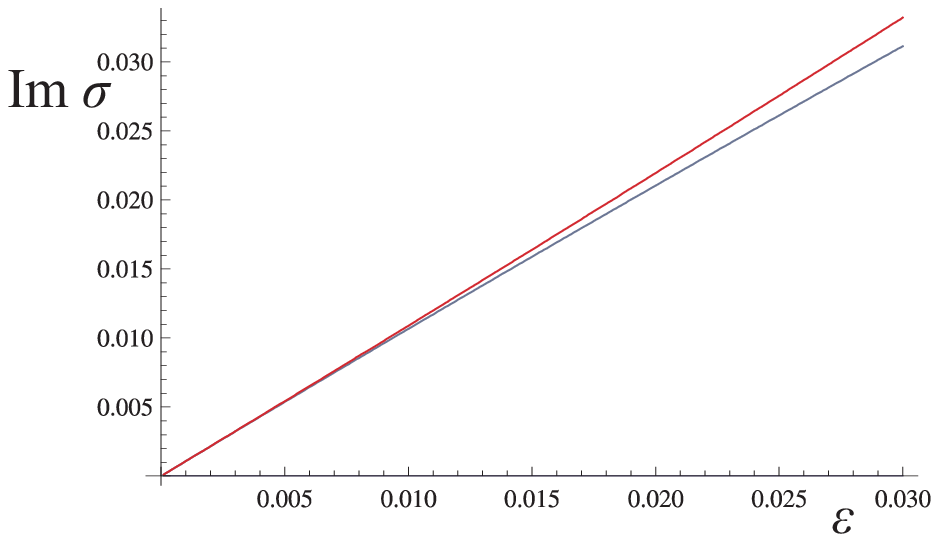}
	\caption{(Left frame) The real part of the characteristic exponents $\sigma_i$ for the charge $S=1$ vortex honeycomb configuration for increasing values of $\ep$. The solid lines represent the values obtained numerically, while the
dashed ones represent the ${\cal O}(\sqrt{\ep})$ theoretical prediction. (Right frame) The imaginary part of the exponents with the double real part.}
	\label{fig:eig_hon_s1}
\end{figure}

We consider now the six-site vortex configuration with $\phi_i=\phi=\pi/3$,
i.e., with $S=1$. As discussed in section \ref{s1}, this configuration is
unstable. Indeed, the numerical simulation shows that all the characteristic
exponents of the central oscillators have nonzero real part, as predicted by
the ${\cal O}(\sqrt{\ep})$ estimation of equation ~(\ref{s_hon_in}) [cf. also
with the
DNLS case of \cite{lawetal}]. These
exponents also possess a nonzero imaginary part which is due to higher-order contributions
of the approximation of $\sigma_i$. The calculated value of their real
part with respect to $\ep$ is shown in the left frame of figure ~\ref{fig:eig_hon_s1}, together with the theoretical estimate.
Note that the numerical (solid) line and the theoretical (dashed) line diverge less than in the two previous cases. In addition in the right frame of figure ~\ref{fig:eig_hon_s1} the positive imaginary part of the exponents with the double real part is shown.
Interestingly, many of these findings (presence of imaginary parts in the
next order and higher quality of agreement of the real part of the first-order predictions) are also present in the
DNLS equation for these solutions \cite{lawetal} (cf. also the square-lattice case in \cite{peli_2d}).

\subsubsection{ Charge $S=2$, 6-site vortex breather.}
\begin{figure}[!ht]
	\centering
		\includegraphics{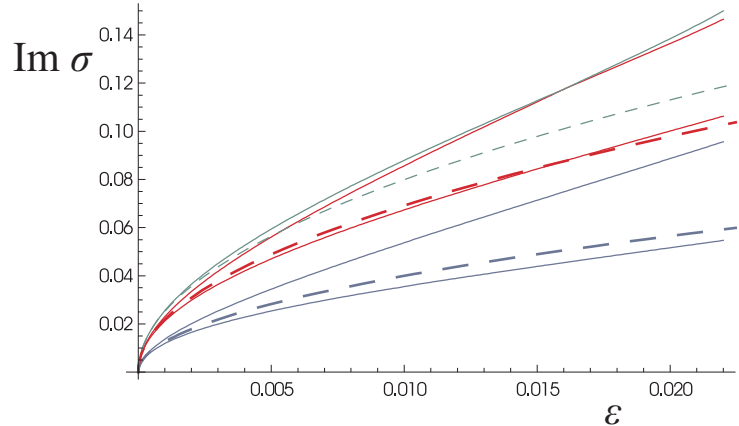}
	\caption{The imaginary part of the characteristic exponents $\sigma_i$ for the charge $S=2$ vortex breather
 honeycomb configuration for increasing values of $\ep$. The solid line represents the values obtained numerically, while the dashed ones represent the ${\cal O}(\sqrt{\ep})$ theoretical prediction.}
	\label{fig:eig_hon_s2}
\end{figure}

The configuration with $\phi_i=\phi=2\pi/3$, i.e., the $S=2$ vortex breather
is linearly stable, as it has been already discussed in section \ref{s2}. Indeed, the numerical simulation confirmed
that the characteristic exponents of the central oscillators are all imaginary and their value with respect to the
value of $\ep$ is shown in figure \ref{fig:eig_hon_s2}, together with the theoretical prediction. Notice that in this case
the higher-order splits the double eigenvalue pairs, similarly to
the case
of the DNLS setting
\cite{lawetal}.

\subsection{Hexagonal lattice}


\subsubsection{In-phase, 3-site breather.}
\begin{figure}[!ht]
	\centering
		\includegraphics{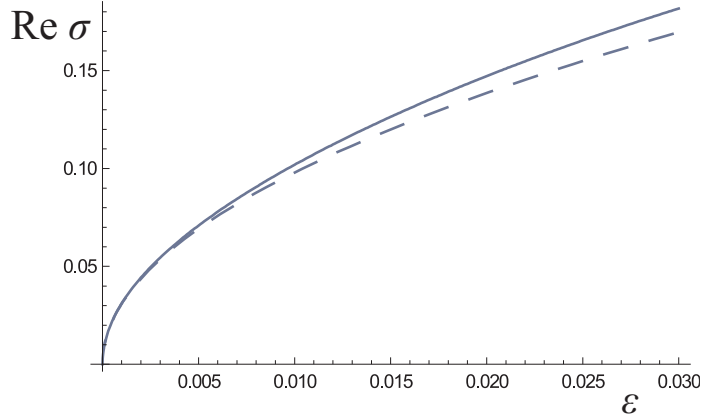}
	\caption{The real part of the characteristic exponents $\sigma_i$ for the in-phase three-site hexagonal configuration is shown
for increasing values of $\ep$. The solid line represents the values
obtained numerically, while the dashed ones
represent the ${\cal O}(\sqrt{\ep})$ theoretical prediction. Notice that the double
pair does not appear to split for the couplings considered.}
	\label{fig:eig_hex_in_re}
\end{figure}

First, we consider the in-phase configuration ($\phi_i=\phi=0$). The corresponding characteristic exponents are shown in
figure ~\ref{fig:eig_hex_in_re}. The solid line represents the value of the exponent which is acquired from the numerical simulation,
while the dashed one represents the theoretical prediction acquired from equation ~(\ref{s_in}).
Note once again in this case the coincidence (also to higher-orders) of the double pair of exponents.
%

\subsubsection{Out-of-phase, 3-site breather.}
\begin{figure}[!ht]
	\centering
		 \includegraphics[width=6cm]{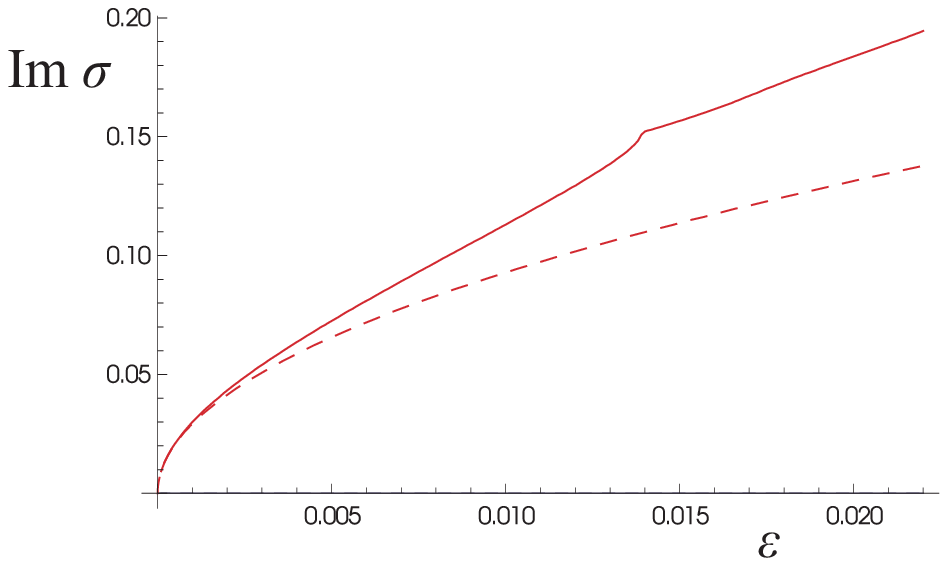} \ \ \includegraphics[width=6cm]{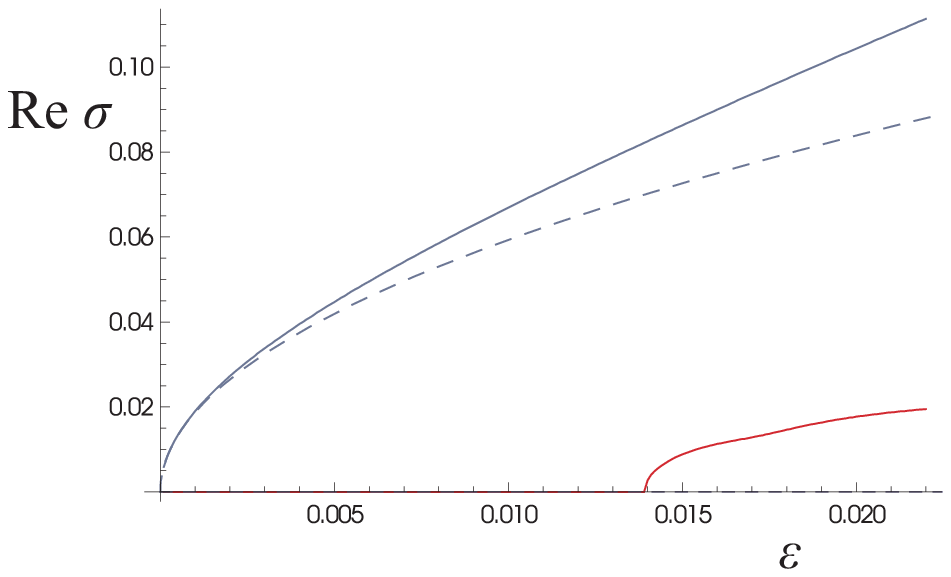}
	\caption{The imaginary (a) and the real part (b) of the characteristic exponents $\sigma_i$
for the out-of-phase three-site hexagonal configuration for increasing values of $\ep$. The solid lines represent the values obtained numerically,
while the dashed ones represent the ${\cal O}(\sqrt{\ep})$ theoretical prediction. Notice the Hamiltonian Hopf bifurcation at
$\ep \simeq 0.014$, leading to the formation of a quartet.}
	\label{fig:eig_hex_out}
\end{figure}

The next configuration under consideration is the out-of-phase one
($\phi_i=\phi=\pi$). This is a linearly unstable configuration, since,
as
seen from equation ~(\ref{s_out}), one of the characteristic exponents
will be real while the other will be imaginary. This is shown also in
panels (a) and (b) of figure ~\ref{fig:eig_hex_out}, where
the imaginary and real parts
of the exponents are respectively shown.
It is observed that, for small values of $\ep$, one of the exponents is purely imaginary while the other is real.
When $\ep$ acquires the value $\ep\simeq0.014$ a Hamiltonian-Hopf bifurcation occurs, which forms a complex quadruplet of exponents
and the corresponding exponent acquires a nonzero real part (this collision with a
mode of the continuous spectrum leads also to the apparent non-smoothness of the line in panel (a)).

\subsubsection{Vortex 3-site breather.}
\begin{figure}[!ht]
	\centering
		\includegraphics{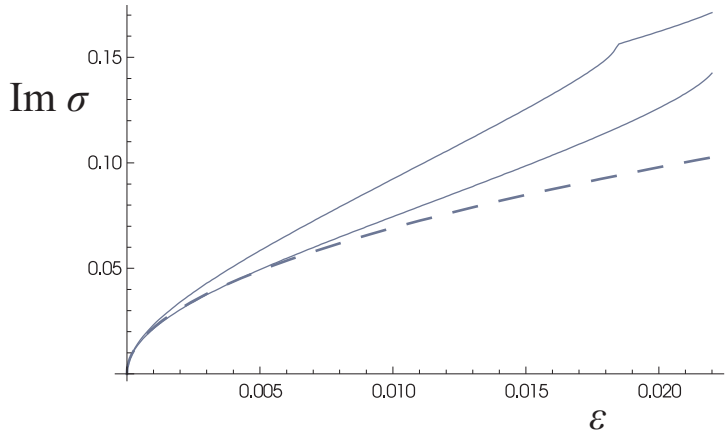}
	\caption{The imaginary part of the characteristic exponents $\sigma_i$ for the vortex three-site hexagonal configuration for increasing values of $\ep$. The solid lines represent the values obtained numerically, while the
dashed ones represent the ${\cal O}(\sqrt{\ep})$ theoretical prediction. Notice
the numerical splitting of the double pair and the Hamiltonian-Hopf bifurcation inducing collision for $\ep \simeq 0.019$.}
	\label{fig:eig_hex_vortex}
\end{figure}

The last configuration under consideration is the vortex
of topological charge $S=1$ ($\phi_i=\phi=2\pi/3$).
This configuration is a stable one as
it can be seen from equation (\ref{s_vortex}). In figure \ref{fig:eig_hex_vortex} the value of the characteristic exponents,
which are both imaginary, are shown. We see in this case that the two branches
corresponding to the numerical acquired values of the exponents are distinct, although we have only one theoretical prediction (i.e., the double
pair splits). This fact implies that the higher-order terms contributing to these exponents are different.
Notice also for $\ep \simeq 0.019$,
a collision of one of the pairs with the linear spectrum, leading to the
emergence of an eigenvalue quartet through a Hamiltonian-Hopf
bifurcation. Note that we
do not consider the topological charge $S=2$ ($\phi_i=\phi=4\pi/3$) configuration in our study. This happens because the $S=2$ case does not produce any physically distinct motion, instead it provides the same motion as the $S=1$ case with the reverse rotation direction.

\subsubsection{In-phase 6-site breather.}
\begin{figure}[!ht]
	\centering
		\includegraphics{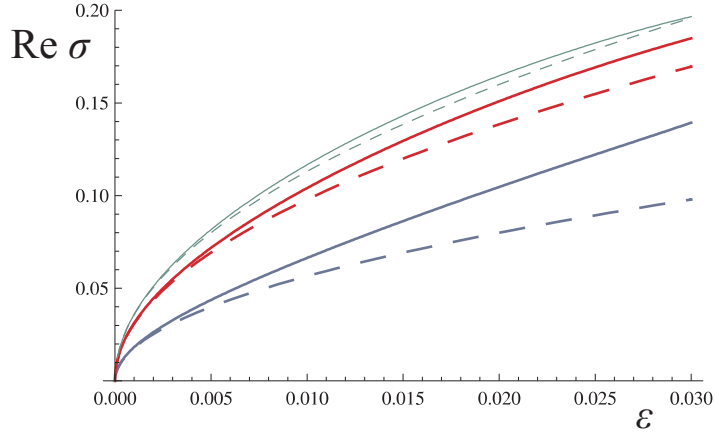}
	\caption{The real part of the characteristic exponents $\sigma_i$ for the in-phase six-site hexagonal
configuration for increasing values of $\ep$. The solid lines represent the values obtained numerically, while the
dashed ones represent the ${\cal O}(\sqrt{\ep})$ theoretical prediction.}
	\label{fig:eig_hex_big_in}
\end{figure}

We consider now $6$-site breathers. The first configuration we study is the in-phase one $(\phi_i=\phi=0)$. As it is already predicted this configuration is unstable. The corresponding characteristic exponents are shown in
figure ~\ref{fig:eig_hex_big_in}. The solid line represents the value of the exponent which is acquired from the numerical simulation,
while the dashed one represents the theoretical prediction acquired from equation (\ref{s_hon_in}).
Note once again in this case the coincidence (also to higher-orders)
of the exponents which have been predicted to be double in the leading
order of approximation. Although the theoretical estimation of the
exponents is the same as in the in-phase honeycomb configuration,
their actual value
 is different because of the different behaviour of the higher order terms.

\subsubsection{Out-of-phase 6-site breather.}
\begin{figure}[!ht]
	\centering
		\includegraphics{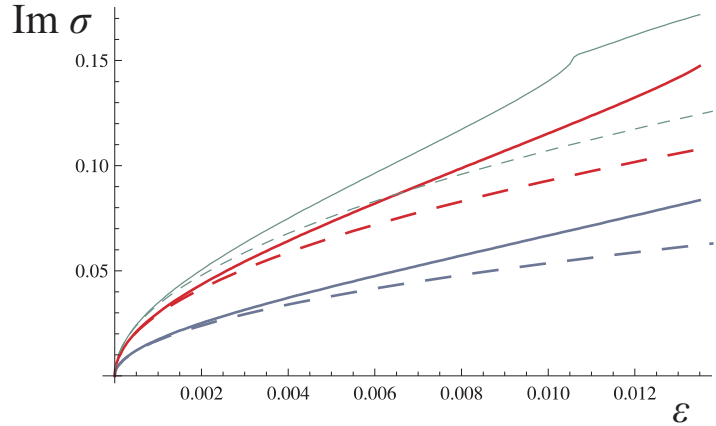}
	\caption{The imaginary part of the characteristic exponents $\sigma_i$ for the out-of-phase six-site hexagonal configuration for increasing values of $\ep$. The solid lines represent the values obtained numerically, while the
dashed ones represent the ${\cal O}(\sqrt{\ep})$ theoretical prediction. Notice
the numerical splitting of the double pair and the Hamiltonian-Hopf bifurcation inducing collision for $\ep \simeq 0.011$.}
	\label{fig:eig_hex_big_out}
\end{figure}

The next configuration under consideration is the out-of phase one
($\phi_i=\phi=\pi$), which is linearly stable for small values of
$\ep$. Indeed as it shown in figure ~\ref{fig:eig_hex_big_out} all the exponents are purely imaginary until $\ep=0.011$ where the Hamiltonian Hopf bifurcation occurs.

\subsubsection{$S=1$, 6-site vortex breather.}
\begin{figure}[!ht]
	\centering
		\includegraphics[width=6cm]{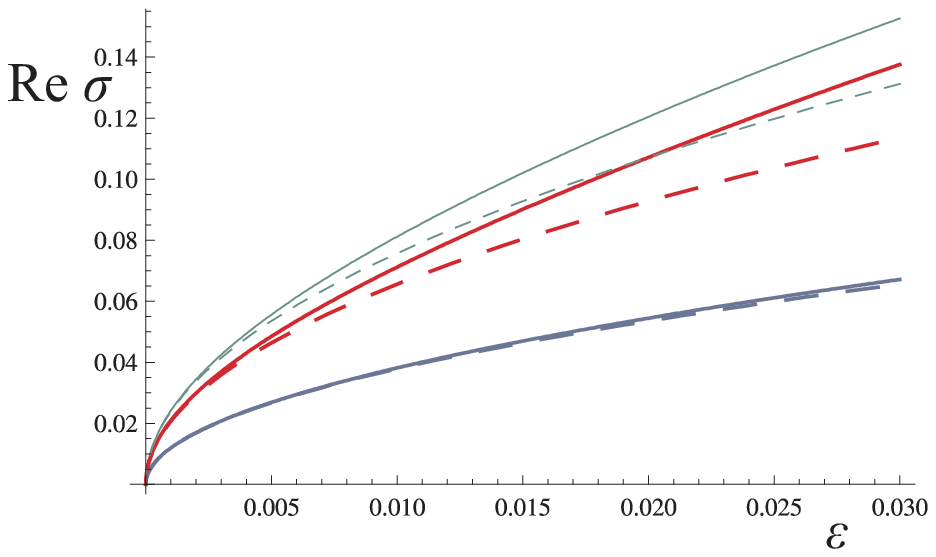} \ \ \includegraphics[width=6cm]{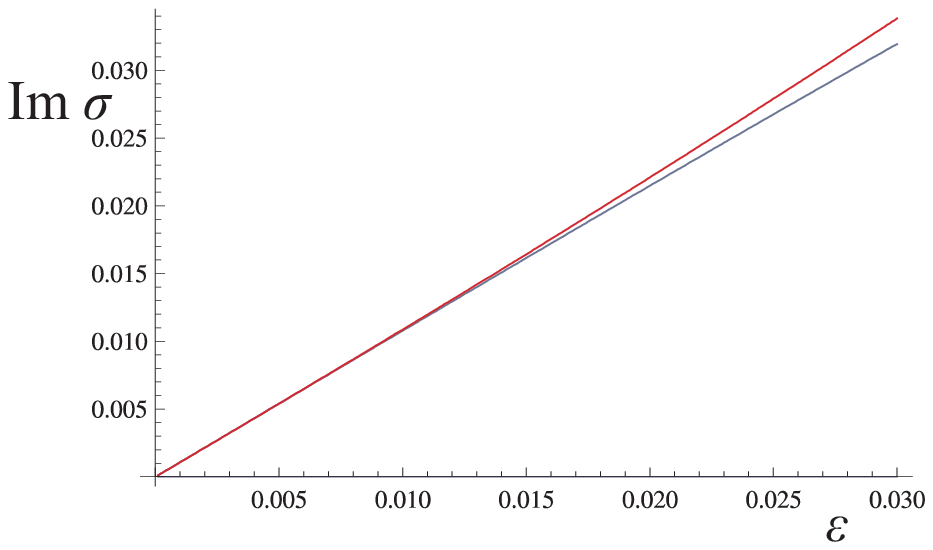}
	\caption{(Left frame) The real part of the characteristic exponents $\sigma_i$ for the $S=1$ vortex six-site hexagonal configuration for increasing values of $\ep$. The solid lines represent the values obtained numerically, while the
dashed ones represent the ${\cal O}(\sqrt{\ep})$ theoretical prediction. (Right frame) the imaginary part of the exponents with the double real part.}
	\label{fig:eig_hex_big_s1}
\end{figure}

We consider now the charge $S=1$ vortex configuration $(\phi_i=\phi=\pi/3)$, which it is anticipated to be linearly unstable, since the theoretical prediction (\ref{s_hon_in}) implies that all the exponents will be real in ${\cal O}(\sqrt{\ep})$. As in corresponding case in the honeycomb lattice, the symplectic signature arguments do not hold and the exponents, which are predicted to have double real part in leading order of approximation, have also a nonzero imaginary part and split, forming two complex quartets, for $\ep$ arbitrary small. This behaviour is shown in figure ~\ref{fig:eig_hex_big_s1}. In the left frame the real part of the exponents $\sigma_i$ is shown together with the ${\cal O(\sqrt{\ep})}$. In the right frame the imaginary part of the exponents is shown. Note that, there is no theoretical prediction for the imaginary part of the exponents, since this is a higher order effect.

\subsubsection{$S=2$, 6-site vortex breather.}
The last configuration under consideration is the $S=2$ vortex breather ($\phi_i=\phi=2\pi/3$) which
is linearly stable. Indeed, as is shown in figure \ref{fig:eig_hon_s2}, the numerical simulation confirmed
that the characteristic exponents of the central oscillators are all
imaginary; their dependence
on $\ep$ is depicted in the figure,
together with the corresponding
theoretical prediction. Notice that, in this case,
the higher-orders of the characteristic exponents development causes
the
splitting of those which have double imaginary part up to leading order of approximation.

\begin{figure}[!ht]
	\centering
		\includegraphics{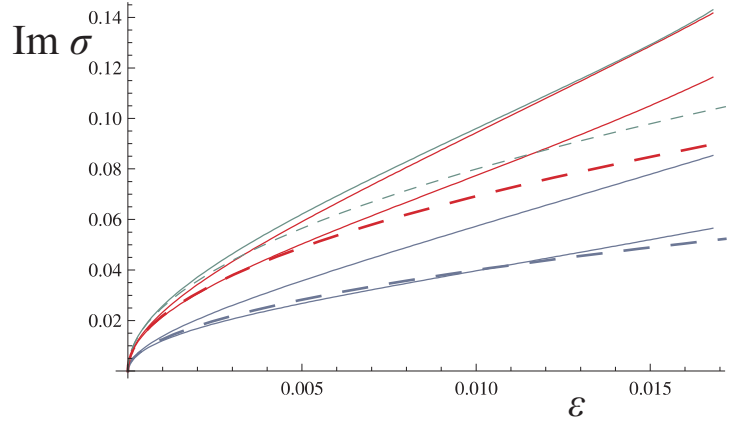}
	\caption{The imaginary part of the characteristic exponents $\sigma_i$ for the $S=2$ vortex six-site hexagonal configuration for increasing values of $\ep$. The solid lines represent the values obtained numerically, while the
dashed ones represent the ${\cal O}(\sqrt{\ep})$ theoretical prediction.}
	\label{fig:eig_hex_big_s2}
\end{figure}

\section{Conclusions}
We have studied the existence and stability of multisite discrete breathers
in two prototypical examples of non-square two-dimensional lattices. The
lattice types under consideration were a honeycomb and a hexagonal
one. We have considered six-site contours in the honeycomb lattice and three-site as well as six-site
ones in the hexagonal case. We have categorized these solutions in terms of them having zero or nonzero energy flux between their bonds, obtaining,
respectively, regular discrete breather as well as vortex breather structures.

Our analysis has offered a systematic analysis of the linear stability of the discrete excitations supported in the prototypical lattices studied.
In the honeycomb lattice case we have considered six-site
configurations and have shown that, for soft potential and positive coupling, the
out-of-phase breather configuration and the charge-two vortex
breather are linearly stable, while the in-phase and charge-one vortex states
are unstable.
In a hexagonal lattice, and in the case of
a soft potential and positive coupling, the in-phase three-site configuration
is unstable and the charge-one vortex is linearly stable.  The out-of-phase three-site configuration here is always unstable, while
the stability results for all other configurations in both lattice cases are reversed in the
setting of either hard potential or negative coupling (but not
both). {The stability results in the six-site hexagonal
case coincide with the ones acquired in the honeycomb lattice configuration.}

Our study was complemented by numerical computations which have been
shown to be in very good agreement with theoretical predictions.
In order to consider larger amplitude multi-breather configurations, we would need a
higher-order theory which is a natural subject for future investigation. This
higher-order treatment could lead us also to an improved approximation in the
calculation of the characteristic exponents considered herein.

Our results are of relevance in 2D discrete systems characterized by a nonlinear substrate potential, namely including molecular chains as well as dusty plasma (Debye) lattices.
Any particular form/type of on-site potential can fit in our description. Finally, we stress that the scope of our formulation includes inverse-dispersive (backward wave supporting) systems, such as transverse oscillations in dusty plasma crystals.

\ack VK would like to thank the Department of Mathematics and
Statistics of UMASS for its warm hospitality. PGK gratefully acknowledges
support from NSF-DMS-0349023 and NSF-DMS-0806762, as well as from
the Alexander von Humboldt Foundation. I.K.
acknowledges support from a EPSRC-UK Science and Innovation award
to Centre for Plasma Physics at Queen's University Belfast (CPP Grant No.
EP/D06337X/1).


\vspace{5mm}

\begin{thebibliography}{99}


\bibitem{peyrard} M. Peyrard, Nonlinearity {\bf 17}, R1 (2004).


\bibitem{photon} Yu. S. Kivshar and G. P. Agrawal,
{\it Optical solitons: from fibers to photonic crystals}
(Academic Press, San Diego, 2003).


\bibitem{moti3} J.W. Fleischer,
G. Bartal, O. Cohen, T. Schwartz, O. Manela, B. Freedman, M. Segev,
H. Buljan, and N. K. Efremidis,
Opt. Express {\bf 13}, 1780
(2005).

\bibitem{review_opt} D. N.\ Christodoulides,
F.\ Lederer, and Y.\ Silberberg,
Nature \textbf{424},
817
(2003).

\bibitem{sievers} M. Sato, B.~E. Hubbard, and A.~J. Sievers,
Rev. Mod. Phys. {\bf 78}, 137 (2006).

\bibitem{morsch1} O. Morsch and M. Oberthaler,
Rev. Mod. Phys. {\bf 78}, 179 (2006);
V.A. Brazhnyi and V.V. Konotop, Mod. Phys. Lett. B
{\bf 18}, 627 (2004).

\bibitem{sen08}
S. Sen, J. Hong, J. Bang, E. Avalos, and R. Doney,
Phys. Rep. {\bf 462}, 21 (2008).

\bibitem{dip} J. Yang, I. Makasyuk, A. Bezryadina, and Z. Chen,
Opt. Lett. \textbf{29}, 1662
(2004).

\bibitem{neck} J. Yang,
I. Makasyuk, P. G. Kevrekidis, H. Martin, B. A. Malomed, D. J.
Frantzeskakis, and Z. Chen,
Phys. Rev. Lett. \textbf{94}, 113902 (2005).

\bibitem{vortex1} D. N. Neshev, T. J. Alexander, E. A. Ostrovskaya, Yu. S. Kivshar,
H. Martin, I. Makasyuk, and Z. Chen,
Phys. Rev. Lett. {\bf 92,} 123903 (2004).

\bibitem{vortex2} J. W. Fleischer, G. Bartal, O. Cohen, O. Manela, M. Segev,
J. Hudock, and D. N. Christodoulides,
Phys. Rev. Lett. \textbf{92}, 123904 (2004).

\bibitem{kartashov} Y. V. Kartashov, V. A. Vysloukh, and L. Torner,
Phys. Rev. Lett. {\bf 93}, 093904 (2004).

\bibitem{rings} X. Wang, Z. Chen, and P. G. Kevrekidis,
Phys. Rev. Lett. \textbf{96}, 083904 (2006).

\bibitem{neshev2} D. Tr{\"a}ger, R. Fischer, D. N. Neshev, A. A. Sukhorukov,
C. Denz, W. Kr{\'o}likowski, and Yu. S. Kivshar,
Optics Express {\bf 14}, 1913
(2006).

\bibitem{motihigher} G. Bartal, O. Manela, O. Cohen, J. W. Fleischer
and M. Segev,
Phys. Rev. Lett. {\bf 95}, 053904 (2005).

\bibitem{zener} H. Trompeter, W. Kr{\'o}likowski, D. N. Neshev,
A. S. Desyatnikov, A. A. Sukhorukov, Yu. S. Kivshar, T. Pertsch, U.
Peschel, and F. Lederer,
Phys. Rev. Lett. {\bf 96}, 053903 (2006).

\bibitem{honey} O. Peleg, G. Bartal. B. Freedman, O. Manela,
M. Segev, and D. N. Christodoulides,
Phys. Rev. Lett. {\bf 98}, 103901 (2007).

\bibitem{rosberg2} C. R. Rosberg, D. N. Neshev, A. A. Sukhorukov,
W. Kr{\'o}likowski, and Yu. S. Kivshar,
Opt. Lett. {\bf 32}, 397
(2007).

\bibitem{gaid} P. G. Kevrekidis, B. A. Malomed, and Yu. B. Gaididei,
Phys. Rev. E {\bf 66}, 016609 (2002).

\bibitem{tja2} T. J. Alexander, A. S. Desyatnikov, and Yu. S. Kivshar,
Opt. Lett. {\bf 32}, 1293
(2007).


\bibitem{motinature1} B. Freedman, G. Bartal, M. Segev, R. Lifshitz,
D. N. Christodoulides, and J. W. Fleischer,
Nature {\bf 440}, 1166
(2006).

\bibitem{motinature2} T. Schwartz, G. Bartal, S. Fishman, and M. Segev,
Nature {\bf 446}, 52
(2007).

\bibitem{szameit} A. Szameit, Y. V. Kartashov, F. Dreisow, M. Heinrich,
V. A. Vysloukh, T. Pertsch, S. Nolte, A. Tunnermann, F. Lederer, and L. Torner,
Opt. Lett. {\bf 33}, 633
(2008).


\bibitem{klaus}
C. Becker, P. Soltan-Panahi, J. Kronjager, S. Stellmer, K. Bongs, and K. Sengstock,
``Spinor BEC in Triangular Optical Lattices'' in the Proceedings of CLEOE-IQEC (2007).

\bibitem{yannis} V. Koukouloyannis and I. Kourakis,
Phys. Rev. E {\bf 76}, 016402 (2007).


\bibitem{koukkour2} V. Koukouloyannis and I. Kourakis, arXiv:0903.1939.

\bibitem{Vlad1} S. V. Vladimirov, P. V. Shevchenko, and N. F. Cramer,
Phys. Rev. E \textbf{56}, R74 (1997).

\bibitem{SVbook} S. V. Vladimirov, K. Ostrikov, A. A. Samarian,
\textit{Physics and Applications of Complex Plasmas} (Imperial
College Press, London, 2005).


\bibitem{IKIJBC}  I. Kourakis and P. K. Shukla, Int. J. Bifurcation
Chaos, \textbf{16} (6), 1711 (2006).

\bibitem{PKSbook} P.K. Shukla and A. A. Mamun, Introduction to Dusty Plasma Physics, IoP Publishing LtD (London, 2002).

\bibitem{GEMreview} G. E. Morfill and A. V. Ivlev, Rev. Mod. Phys. \textbf{81}, 1353 (2009).

\bibitem{lawetal} K. J. H. Law, P. G. Kevrekidis, V. Koukouloyannis,
I. Kourakis, D. J. Frantzeskakis, and A.R. Bishop,
Phys. Rev. E {\bf 78}, 066610 (2008).

\bibitem{koukmac} V. Koukouloyannis and R. S. MacKay,
J. Phys. A: Math. Gen. {\bf 38}, 1021 (2005).


\bibitem{alexlaw} K. J. Law, P. G. Kevrekidis, T. J. Alexander,
W. Kr{\'o}likowski, and Yu. S. Kivshar,
Phys. Rev. A {\bf 79}, 025801 (2009).

\bibitem{berndlaw} B. Terhalle, T. Richter, K. J. H. Law,
D. G{\"o}ries, P. Rose, T. J. Alexander, P. G. Kevrekidis, A. S.
Desyatnikov, W. Kr{\'o}likowski, F. Kaiser, C. Denz, and Yu. S. Kivshar,
Phys. Rev. A {\bf 79}, 043821 (2009).

\bibitem{macaub} R. S. MacKay and S. Aubry, Nonlinearity {\bf 7}, 1623 (1994).

\bibitem{ahnmacsep} T. Ahn, R. S. MacKay, and J. A. Sepulchre, Nonlinear Dynamics {\bf 25}, 157 (2001).

\bibitem{koukicht1} V. Koukouloyannis and S. Ichtiaroglou,
Phys. Rev. E {\bf 66}, 066602 (2002).

\bibitem{creaub} T. Cretegny and S. Aubry, Physica D {\bf 113}, 162 (1998).

\bibitem{video_link} http://users.auth.gr/vkouk/hon-hex-videos.zip


\bibitem{aubrev} S. Aubry, Physica D {\bf 103}, 201 (1997).

\bibitem{mac2} R. S. MacKay, in {\it Hamiltonian Dynamical Systems},
edited by R. S. MacKay and J. D. Meiss (Adam Hilger, 1987), pp. 137.


\bibitem{peli_2d} D. E. Pelinovsky, P. G. Kevrekidis, and D. J.
Frantzeskakis, Physica D {\bf 212}, 20 (2005).

\end{thebibliography}
\end{document}